\documentclass[10pt,twocolumn,superscriptaddress,longbibliography,prx]{revtex4-2}

\usepackage{amsmath, bm, amsfonts, amssymb}     
\usepackage{graphicx}                       
\usepackage{xfrac}                          
\usepackage{grffile}                        
\usepackage{xcolor}

\newcommand{\gdot}{\dot{\gamma}}
\newcommand{\tw}{t_{\rm w}}

\newcommand{\dobmean}{\langle{\Delta\gdot}\rangle}

\usepackage[colorlinks, citecolor=blue, urlcolor=red, linkcolor=blue]{hyperref}

\begin{document}

\title{Slow fatigue and highly delayed yielding via shear banding  in oscillatory shear}
\author{James O. Cochran}
\affiliation{Department of Physics, Durham University, Science Laboratories,
  South Road, Durham DH1 3LE, UK}
 \author{Grace L. Callaghan}
\affiliation{Department of Physics, Durham University, Science Laboratories,
  South Road, Durham DH1 3LE, UK}
 \author{Miles J. G. Caven}
\affiliation{Department of Physics, Durham University, Science Laboratories,
  South Road, Durham DH1 3LE, UK}
\author{Suzanne M. Fielding}
\affiliation{Department of Physics, Durham University, Science Laboratories,
  South Road, Durham DH1 3LE, UK}

\begin{abstract}

We study theoretically the dynamical process of yielding in cyclically sheared amorphous materials, within a thermal elastoplastic model and the soft glassy rheology model. Within both models we find an initially slow accumulation, over many cycles after the inception of shear, of low levels of damage in the form strain heterogeneity across the sample.  This slow fatigue then suddenly gives way to catastrophic yielding and material failure. Strong strain localisation in the form of shear banding is key to the failure mechanism. We characterise in detail the dependence of the number of cycles $N^*$ before failure on the amplitude of imposed strain, the working temperature, and the degree to which the sample is annealed prior to shear.  We discuss our finding with reference to existing experiments and particle simulations, and suggest new  ones to test our predictions.
\end{abstract}

\maketitle

Amorphous materials~\cite{nicolas2018deformation,bonn2017yield,berthier2011theoretical} include soft solids such as emulsions, colloids, gels and granular materials, and harder metallic and molecular glasses. Unlike crystalline solids, they lack order in the arrangement of their constituent microstructures (droplets, grains, {\it etc.}). Understanding their rheological properties is  thus  a major challenge. Typically, they behave elastically at low loads then yield plastically at larger loads. Much effort has been devoted to understanding the  dynamics of yielding following the imposition of a shear stress $\sigma$ or strain rate $\gdot$, which is held constant after switch-on. This often involves the formation of shear bands~\cite{fielding2016triggers}, which  can slowly heal away to leave homogeneous flow in complex fluids~\cite{divoux2010transient,shrivastav2016heterogeneous,vasisht2020computational,vasisht2020emergence,divoux2012yielding1,grenard2014timescales,benzi2019unified,chaudhuri2013onset,manning2007strain1,hinkle2016small1,jagla2007strain}, or trigger catastrophic failure in solids~\cite{hufnagel2016deformation,doyle1972fracture1}. In many  applications, however, materials are subject to a {\em cyclically repeating} deformation or load. Cyclic shear is also important fundamentally, in revealing key fingerprints of a material's nonlinear rheology, with large amplitude oscillatory shear intensely studied~\cite{rouyer2008large,yoshimura1987response,viasnoff2003colloidal,ewoldt2010large,renou2010yielding,guo2011correlations,van2013rheology,koumakis2013complex,poulos2013flow,poulos2015large,rogers2011sequence,rogers2012sequence,de2013unified,blackwell2014simple,radhakrishnan2016shear,radhakrishnan2018shear,van2018strand}.

The response of an amorphous material to an oscillatory shear strain depends strongly on the strain amplitude $\gamma_0$  relative to a threshold $\gamma_{\rm c}$~\cite{bhaumik2021role,corte2008random,das2020unified,fiocco2013oscillatory,kawasaki2016macroscopic,khirallah2021yielding,knowlton2014microscopic,kumar2022mapping,lavrentovich2017period,leishangthem2017yielding,liu2022fate,mungan2021metastability,nagamanasa2014experimental,ness2020absorbing,pine2005chaos,priezjev2018molecular,priezjev2020delayed,regev2013onset,sastry2021models,szulc2020forced,yeh2020glass}. For $\gamma_0<\gamma_{\rm c}$, a material typically settles into deep regions of its energy landscape, showing  reversible response from cycle to cycle (after many cycles), via an absorbing state transition. The number of cycles to settle however diverges as $\gamma_0\to\gamma_{\rm c}^-$. For $\gamma_0>\gamma_{\rm c}$, a material instead yields into a state of higher energy that is chaotically irreversible from cycle to cycle, and often shear banded~\cite{radhakrishnan2016shear,radhakrishnan2018shear,parmar2019strain,yeh2020glass,denisov2015sharp}.

Indeed, the process of repeatedly straining or loading a material over many cycles typically leads to the gradual accumulation of microstructural damage. While the early signatures of such fatigue are often difficult to detect, its slow buildup can eventually undermine material stability and precipitate catastrophic failure. Understanding the  accumulation of microstructural fatigue and identifying the microscopic precursors that prefigure failure is thus central to the prediction of material stability and lifetime, and the development of strategies to improve them. 

In hard materials, the buildup of microstructural damage is often interpreted in terms of the formation of microcracks. Far less well understood in soft materials, it remains the topic of intense study, as recently reviewed~\cite{cipelletti2020microscopic}. Colloidal gels in oscillatory stress~\cite{perge2014time,gibaud2016multiple,saint2017predicting} display an intricate, multi-stage yielding process in which the sample remains  solid-like for many cycles, before slipping at the rheometer wall, then forming coexisting solid-fluid bulk shear bands and finally fully fluidizing~\cite{saint2017predicting}.  The number of cycles before yielding increases dramatically  at low stress amplitudes~\cite{gibaud2016multiple,saint2017predicting}.  Particle~\cite{bhowmik2022fatigue} and fibre bundle~\cite{kun2007fatigue} simulations likewise show increasing yielding delay with decreasing cyclic load amplitude. Metallic glass simulations show an increasing number of cycles to shear band formation with decreasing $\gamma_0$~\cite{sha2015cyclic}. Particle simulations~\cite{corte2008random,fiocco2013oscillatory} and experiments on colloidal glass~\cite{nagamanasa2014experimental} show a number of strain cycles to attain a yielded steady state diverging as $\gamma_0\to\gamma_{\rm c}^+$.

Despite this rapid experimental progress, the dynamics of yielding in cyclic shear remains poorly understood theoretically.  An insightful recent study of athermal materials captured delayed yielding after a number of cycles that increases at low strain amplitude~\cite{parley2022mean}. In being mean field, however, this work necessarily neglects the development of damage in the form of strain heterogeneity and shear bands that are key to understanding yielding.

In this Letter, we study theoretically the yielding of amorphous materials in oscillatory shear strain. Our contributions are fourfold. First, we predict a slow accumulation, over many cycles, of initially low levels of damage in the form of strain heterogeneity across the sample.  Second, we show that this early fatigue later gives way to catastrophic material failure, after a number of cycles $N^*$. Third, we show that the formation of shear bands is key to the failure mechanism, as seen experimentally. Finally, we characterise the dependence of $N^*$ on the strain amplitude $\gamma_0$, the working temperature $T$, and the degree of sample annealing prior to shear. 

{\it Models ---} To gain confidence that our predictions are generic across a wide range of amorphous materials, independent of specific constitutive modelling assumptions, we study numerically two different widely used models of elastoplastic rheology: the soft glassy rheology (SGR) model~\cite{sollich1997rheology} and a thermal elastoplastic (TEP) model~\cite{nicolas2018deformation}. 

The SGR model comprises an ensemble of elastoplastic elements, each corresponding to a mesoscopic region of material large enough to admit a local continuum shear strain $l$ and stress $Gl$, with modulus $G$.  Under an imposed shear rate $\dot\gamma$, any element strains at rate $\dot l = \dot\gamma$. Elemental stresses are however intermittently released via local plastic yielding events, occurring stochastically at rate $r = \tau_0^{-1} {\rm min} \{1,\exp\left[-\left(E - \frac{1}{2}Gl^2\right)/T\right]\}$, with $\tau_0$ a microscopic attempt time, $E$ a local energy barrier, and $T$ temperature. Upon yielding, any element resets its  strain, $l\to 0$, and selects a new yield energy from a distribution $\rho\left( E\right) = \exp\left(-E/T_g \right)/T_g$. The model captures a glass transition at temperature $T=T_g$ and predicts rheological aging at low loads in its glass phase, $T < T_g$.  The macroscopic elastoplastic stress  $\sigma$ is the average of the elemental stresses. The total stress $\Sigma=\sigma+\eta\gdot$ includes a Newtonian contribution of viscosity $\eta$.

The TEP model is defined likewise, except each element has the same yield energy $E$, and after yielding selects its new $l$ from a Gaussian  of width $l_{\rm h}$.  Both models thus combine the basic ingredients of elastic deformation punctuated by plastic rearrangements and stress propagation. But whereas SGR  incorporates disorder in the material's energy landscape via $\rho(E)$ to  capture glassy behaviour, yet neglects frustrated local stresses, the TEP model conversely neglects glassiness, but captures frustrated local stresses via the post-hop $l-$distribution.

To capture catastrophic yielding, it is crucial to allow for strain localisation and shear banding. Accordingly, in each model the elastoplastic elements are arranged across  $S$ streamlines stacked in the flow gradient direction $y$, with $M$ elements per streamline. The imposed shear rate, averaged across streamlines,  is $\bar{\dot\gamma}\left(t\right)$. The local shear rate can however vary across streamlines: imposing uniform total stress $\Sigma$ in creeping flow gives $\Sigma\left(t\right) = \sigma\left(y,t\right) + \eta\dot\gamma\left(y,t\right) = \bar\sigma\left(t\right) + \eta \bar{\dot\gamma}\left(t\right)$, with $y$ a streamline's flow gradient coordinate. After any local yielding event with stress drop of magnitude $\Delta l$ we furthermore pick three random elements on each neighbouring streamline and adjust their $l\to l + w\Delta l \left(-1,+2,-1\right)$. We thus implement 1D Eshelby stress propagation~\cite{picard2004elastic} and stress diffusion~\cite{lu2000effects}, which are key to shear banding.

{\it Protocol ---}  We study oscillatory shear strain $\bar{\gamma}(t)=\gamma_0\sin(\omega t)$, imposed for all times $t>0$. Prior to shear, the sample is prepared via ageing or annealing. Within SGR, we perform a sudden deep quench at time $t=-\tw$ from infinite temperature to a working temperature $T<T_{\rm g}$ in the glass phase, then age the sample for a waiting time $\tw$. Within TEP, we first equilibrate the sample to a temperature $T_0$, then suddenly at time $t=0$ quench to a working temperature $T<T_0$.  Larger $\tw$ (SGR) or smaller $T_0$ (TEP) corresponds to better annealing.

About an initially uniform state, tiny levels of heterogeneity are seeded naturally via $M$ and $S$ being finite. In SGR we also test the effect of adding a small initial perturbation to the well depths $E \to E \left( 1 + \delta\cos{2\pi y}\right)$. That we observe the same physics in both cases shows that our results are robust to small initial randomicity.

In response to the imposed strain, we measure the shear stress $\Sigma(t)$ and report its root mean square   $\Sigma_{\rm RMS}$ over each cycle vs. cycle number $N$. We also define the degree of shear banding $\Delta\gdot(t)$  via the standard deviation  of the strain rate across streamlines, normalised by  $\gdot_0=\gamma_0\omega$, and report its mean over each cycle, $\dobmean(N)$. When this quantity is high, the strain rate profile is significantly shear banded  across the flow gradient direction.

{\it Parameters ---} Both models have as parameters the mean local yield energy $\langle E\rangle$, attempt time $\tau_0$, temperature $T$, number of streamlines $S$, elements per streamline $M$, Newtonian viscosity $\eta$ and stress diffusion $w$. The degree of annealing is prescribed  by the waiting time $\tw$  (SGR) or pre-quench temperature $T_0$ (TEP). The imposed  shear has amplitude $\gamma_0$ and frequency $\omega$.  We choose units $\tau_0=1$, $G=1$, $\langle E\rangle = 1$. We set  $\eta=0.05$, $w = 0.05$, $l_{\rm h} = 0.05, \delta=0.01$, suited to the Newtonian viscosity, stress diffusivity and initial heterogeneity being small. We set the numerical parameters $S=25$, $M=10000$, having checked for robustness to variations in these. For computational efficiency, we set $\omega=0.1$ in SGR, but checked that our findings also hold for $\omega=0.01$. In TEP we set $\omega = 0.01$. We then explore yielding as a function 
of strain amplitude $\gamma_0$, working temperature $T$, and degree of annealing before shear.

\begin{figure}[t]
    \includegraphics[width=8.5cm]{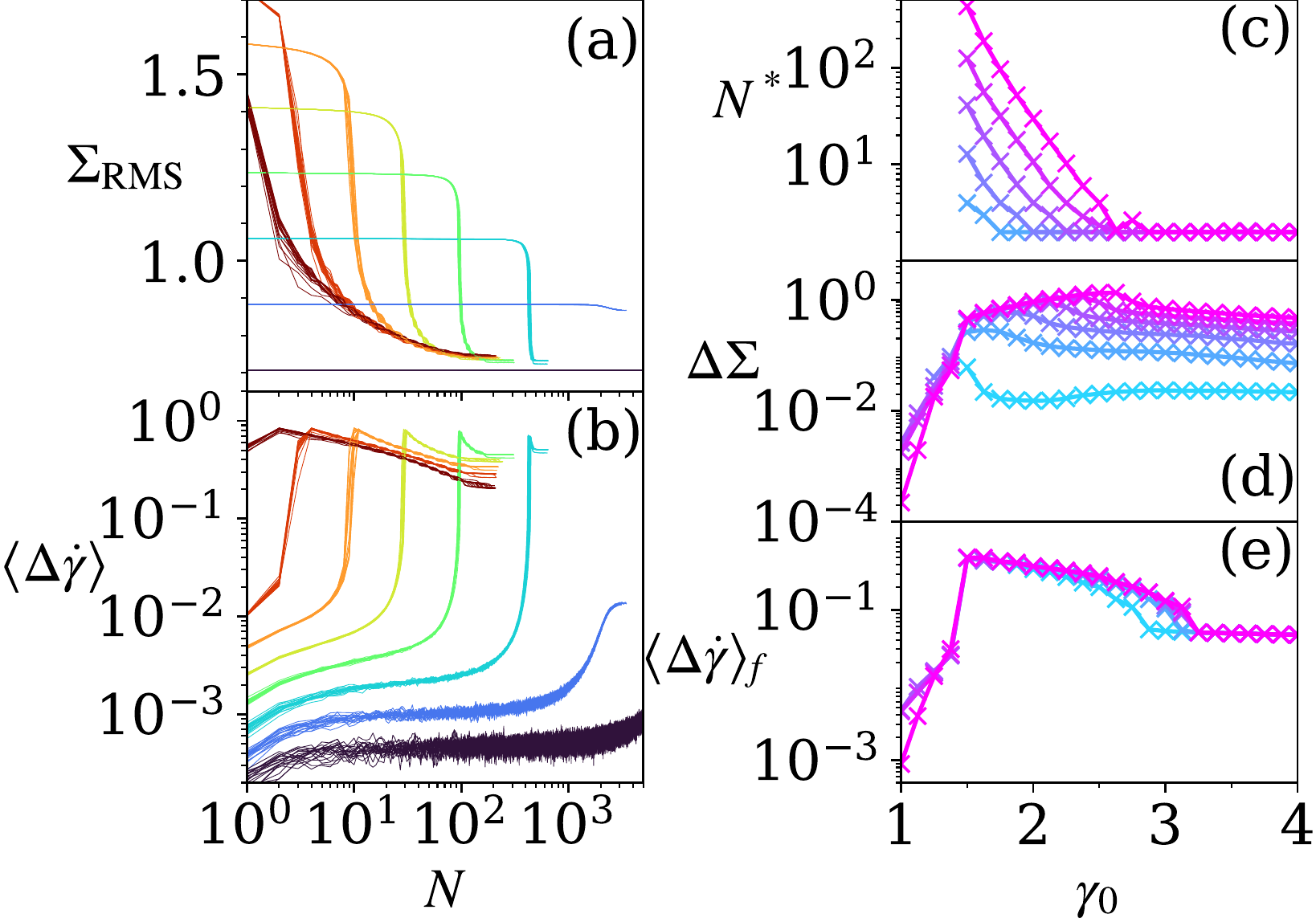}
    \caption{{\bf SGR model.  a)} Root mean square stress and {\bf b)} mean degree of shear banding over each cycle versus cycle number $N$ for strain amplitudes $\gamma_0 = 1.00,1.25,...,2.75$ in curve sets with drops in (a) and rises in (b) right to left.  Each curve within a set corresponds to a different random initial condition. $t_w = 10^7, T=0.3$. {\bf c)} Cycle number at failure $N^*$, {\bf d)} magnitude of stress drop $\Delta\Sigma$ and {\bf e)} final degree of shear banding $\left<\dot\gamma\right>_{\rm f}$ vs. strain amplitude $\gamma_0$ for waiting times $t_w = 10^2, 10^3,...,10^7$  in curves bottom to top. Panel c) only shows samples with $\Delta\Sigma > 0.1$. $N^*,\Delta\Sigma,\left<\dot\gamma\right>_{\rm f}$ averaged over initial condition.
    }
    \label{fig:SGRgamma}
\end{figure}

{\it SGR results ---}  The key physics that we report is exemplified by Figs.~\ref{fig:SGRgamma}a,b). These show that yielding comprised two distinct stages as a function of cycle number $N$. In the first stage, the sample remains nearly homogeneous, with only low level material fatigue (small strain heterogeneity $\dobmean$) slowly accumulating from cycle to cycle, and the stress remaining high.  After a delay that increases dramatically with decreasing imposed strain amplitude $\gamma_0$ in curve sets left to right, a second stage ensues: the stress drops quickly, the strain becomes highly localised into shear bands, and the sample fails catastrophically.

\begin{figure}[t]
    \includegraphics[width=8.6cm]{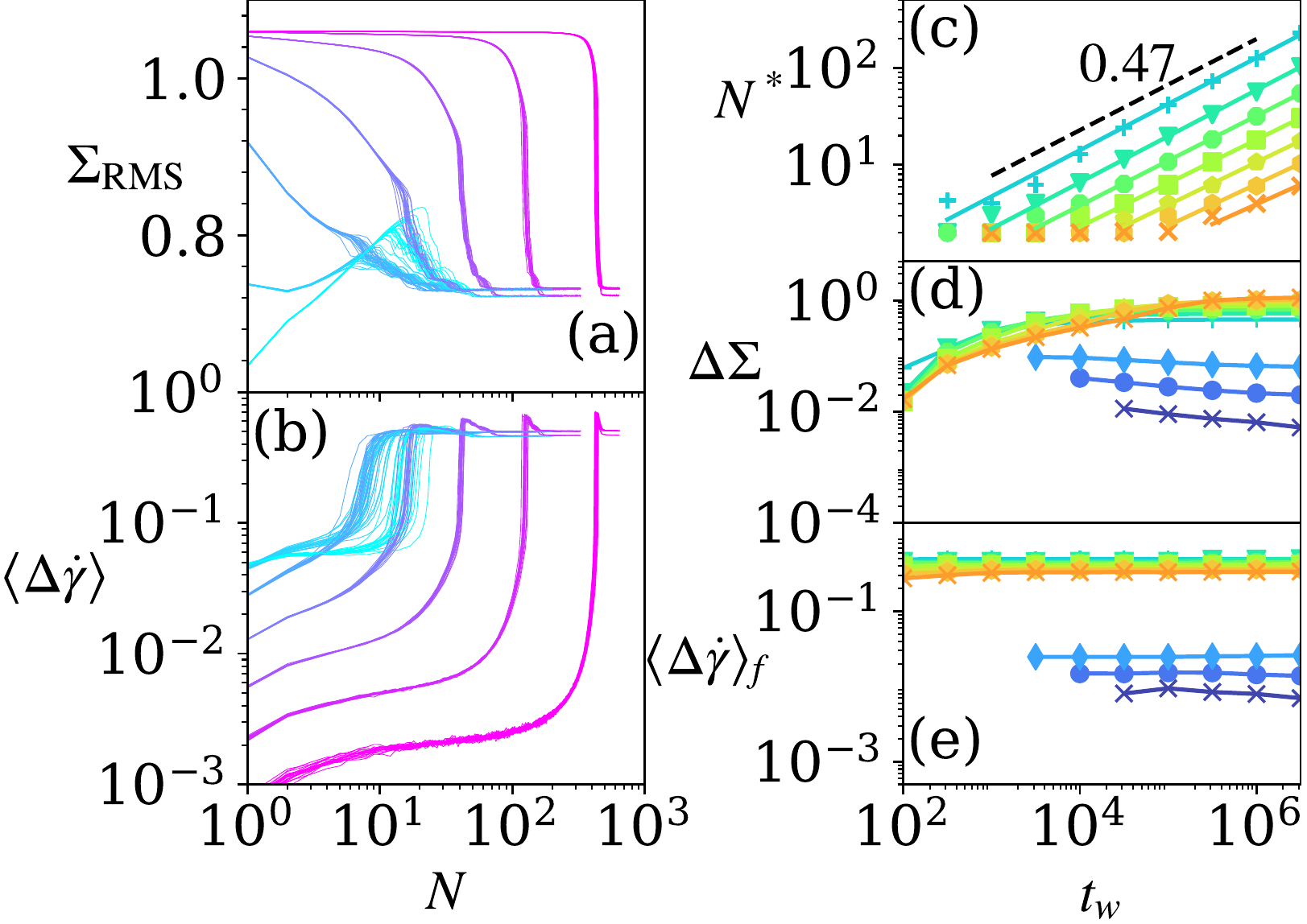}
    \caption{{\bf SGR model. a)} Root mean square stress and {\bf b)} mean  degree of shear banding over each cycle as a function of cycle number $N$ for waiting time $t_w = 10^1, 10^2,...,10^7$ in curve sets with drops in (a) and rises in (b) left to right. 
    $\gamma_0 = 1.5, T=0.3$. {\bf c)} Cycle number at failure $N^*$, {\bf d)} magnitude of stress drop $\Delta\Sigma$, and {\bf e)} final degree of shear banding $\left<\dot\gamma\right>_{\rm f}$ vs.  waiting time, $t_w$. Strain amplitude $\gamma_0 = 1.125,1.250,1.375,...,2.250$ in curves blue to orange, i.e., top to bottom in (c), bottom to top at right of (d), and with $\gamma_0=1.125, 1.25\cdots 1.375$ bottom up and $1.5,...2.25$ top down in (e). Panel c) only shows cases for which $\Delta\Sigma > 0.1$. 
    } 
    \label{fig:SGRtw}
\end{figure}

To quantify the delay during which fatigue slowly accumulates before the sample catastrophically fails, we define the cycle at failure $N^*$ as that in which $\Sigma_{\rm RMS}$ first falls below $\tfrac{1}{2}(\Sigma_{\rm max}-\Sigma_{\rm min})$, where  $\Sigma_{\rm max}$ and $\Sigma_{\rm min}$ are  the global maximum and minimum of $\Sigma_{\rm RMS}$ versus $N$~\footnote{If the minimum occurs before the maximum, we argue that the sample does not show yielding for the parameter values in question.}. We further define the magnitude of yielding via  the normalised stress drop $\Delta\Sigma=(\Sigma_{\rm max}-\Sigma_{\rm min})/\Sigma_{\rm SS}$, where $\Sigma_{\rm SS}$ is the steady state stress as $N\to \infty$; and the extent to which strain becomes localised via the final degree of shear banding $\left<\Delta\gdot\right>_{\rm f}=\lim_{N\to\infty}\left<\Delta\gdot\right>(N)$. These three quantities  are plotted vs. $\gamma_0$ in Figs.~\ref{fig:SGRgamma}c-e).  

Clearly apparent is a transition at strain amplitude $\gamma_0=\gamma_{\rm c}\approx 1.4$, below which the stress drop  $\Delta\Sigma$ and degree of strain localisation $\left<\Delta\gdot\right>_{\rm f}$ become negligible: for $\gamma_0<\gamma_{\rm c}$,  no appreciable yielding occurs. For $\gamma_0>\gamma_{\rm c}$, we see  a range of $\gamma_0$, increasing with increasing $\tw$, over which yielding is both strongly apparent and heavily delayed.  The delay increases dramatically with decreasing $\gamma_0$, although $N^*$ shows no apparent divergence over the window of strains for which yielding is appreciable.

The dependence of yielding on the degree of ageing prior to shear, $\tw$, is further explored in Fig.~\ref{fig:SGRtw}. Panels a) and b) again reveal the two stage yielding just described, with curve sets left to right showing a longer delay with increasing $\tw$, with $N^*\sim\tw^\alpha$ (panel c). Importantly, therefore, ultra annealed samples $\tw\to\infty$ are predicted to show a near indefinite delay before suddenly failing. 

\begin{figure}[t]
    \includegraphics[width=8.5cm]{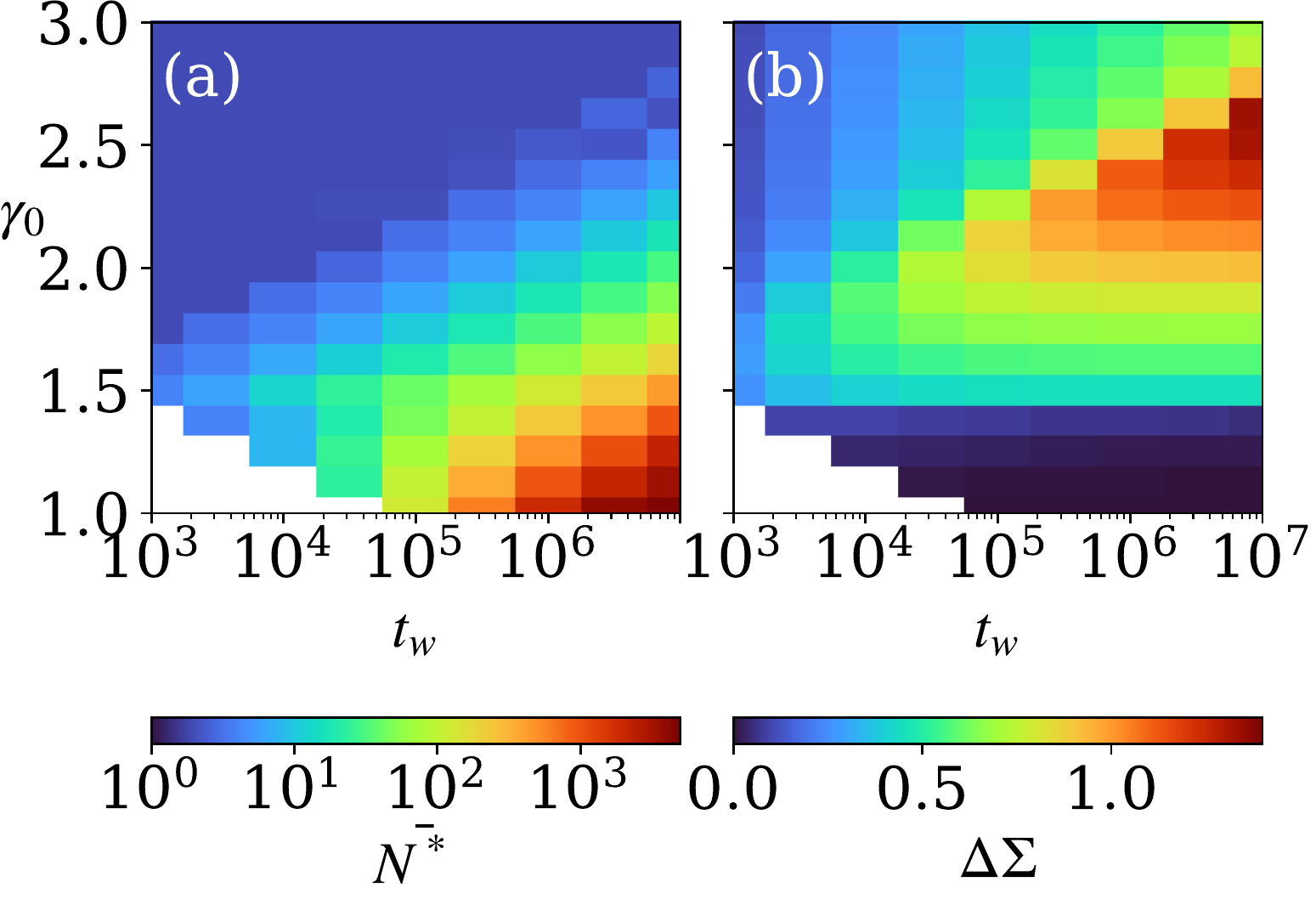}
    \caption{{\bf SGR model} {\bf a)} cycle number at failure $N^*$ and {\bf b)} stress drop $\Delta\Sigma$ as a function of waiting time $t_w$ and strain amplitude $\gamma_0$. In the white region, no yielding occurs.
    }
    \label{fig:phase-space}
\end{figure}

So far, we have characterised the dependence of yielding on the strain amplitude  $\gamma_0$ and waiting time $t_w$ separately. Its dependence on both  parameters is summarised in Fig. \ref{fig:phase-space}.
Importantly, these colormaps suggest the possibility of long delayed (large $N^*$) and catastrophic (large $\Delta\Sigma$) yielding even at large strain amplitudes, provided the sample age prior to shear is large enough. The strain $\gamma_0$ at yielding onset in panel b) roughly coincides with the end of the linear regime, in which the viscoelastic spectra $G'$ and $G''$ are constant functions of $\gamma_0$~\cite{radhakrishnan2018shear}.

\begin{figure}[t]
    \includegraphics[width=8.6cm]{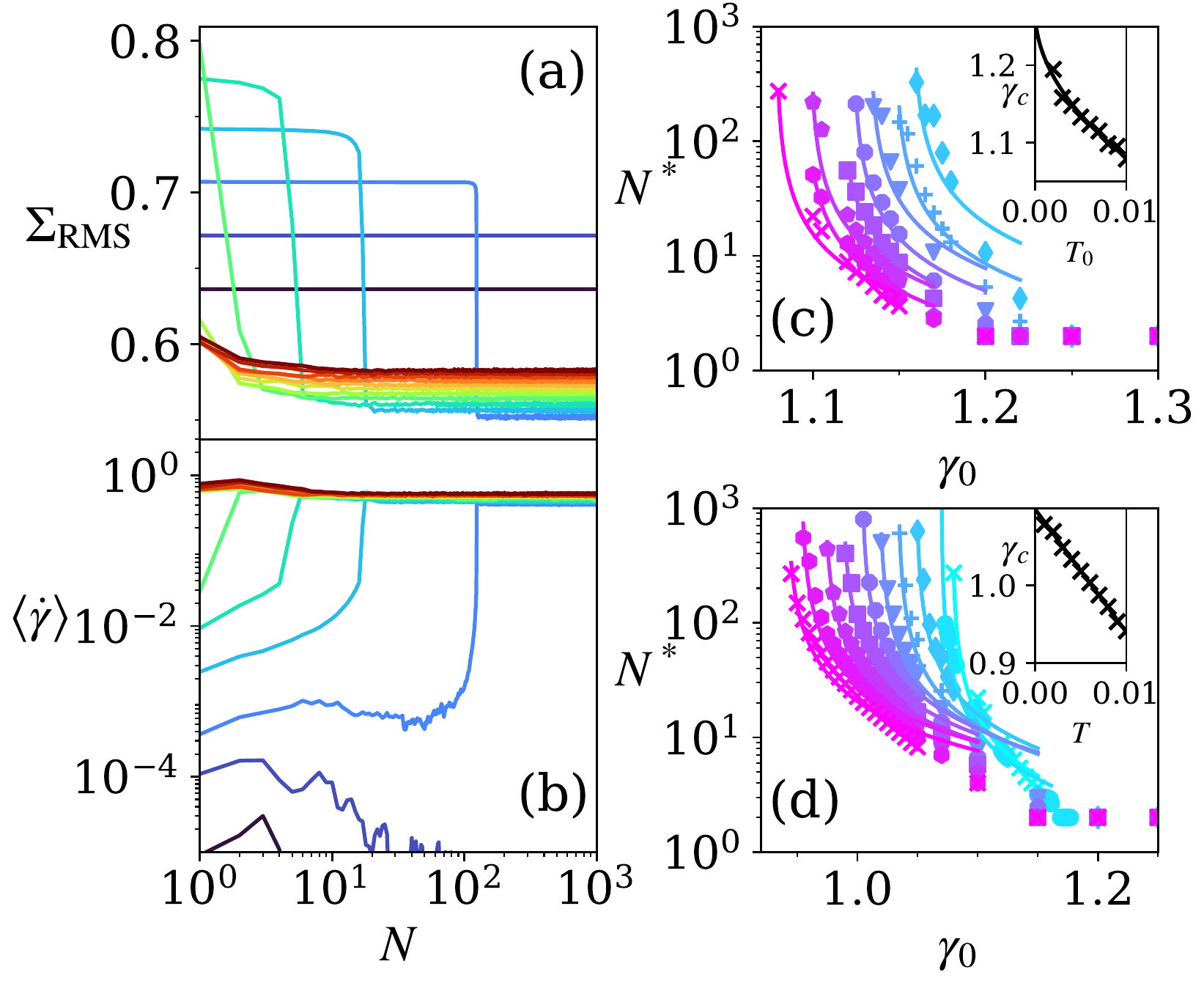}
    \caption{{\bf TEP model.} {\bf  a)} Root mean square stress and {\bf b)} mean degree of shear banding over each cycle as a function of cycle number $N$ for  strain amplitudes $\gamma_0 = 0.90,0.95,...,1.50$ in curve sets with drops in (a) and rises in (b) right to left. $T_0 = 0.01$, $T = 0.007$. Cycle number at yielding $N^*$ vs. strain amplitude $\gamma_0$ for {\bf c)} pre-quench temperatures $T_0 = 0.001,0.002,...,0.010$ in curves right to left at working temperature $T = 0.001$ and {\bf d)} working temperatures $T = 0.001,0.002,...,0.010$ in curves right to left at pre-quench temperature $T_0 = 0.01$. Solid lines in {\bf c)+d)} are fits to $N^* = A/\left(\gamma_0 - \gamma_c\right)$. Insets show $\gamma_c$ (symbols) fit (lines) to {\bf c)} $\gamma_c = B - C\sqrt{T_0}$ and {\bf d)} $\gamma_c = DT - E$.
    }
    \label{fig:TEP1}
\end{figure}

\begin{figure}[t!]
    \includegraphics[width=8.6cm]{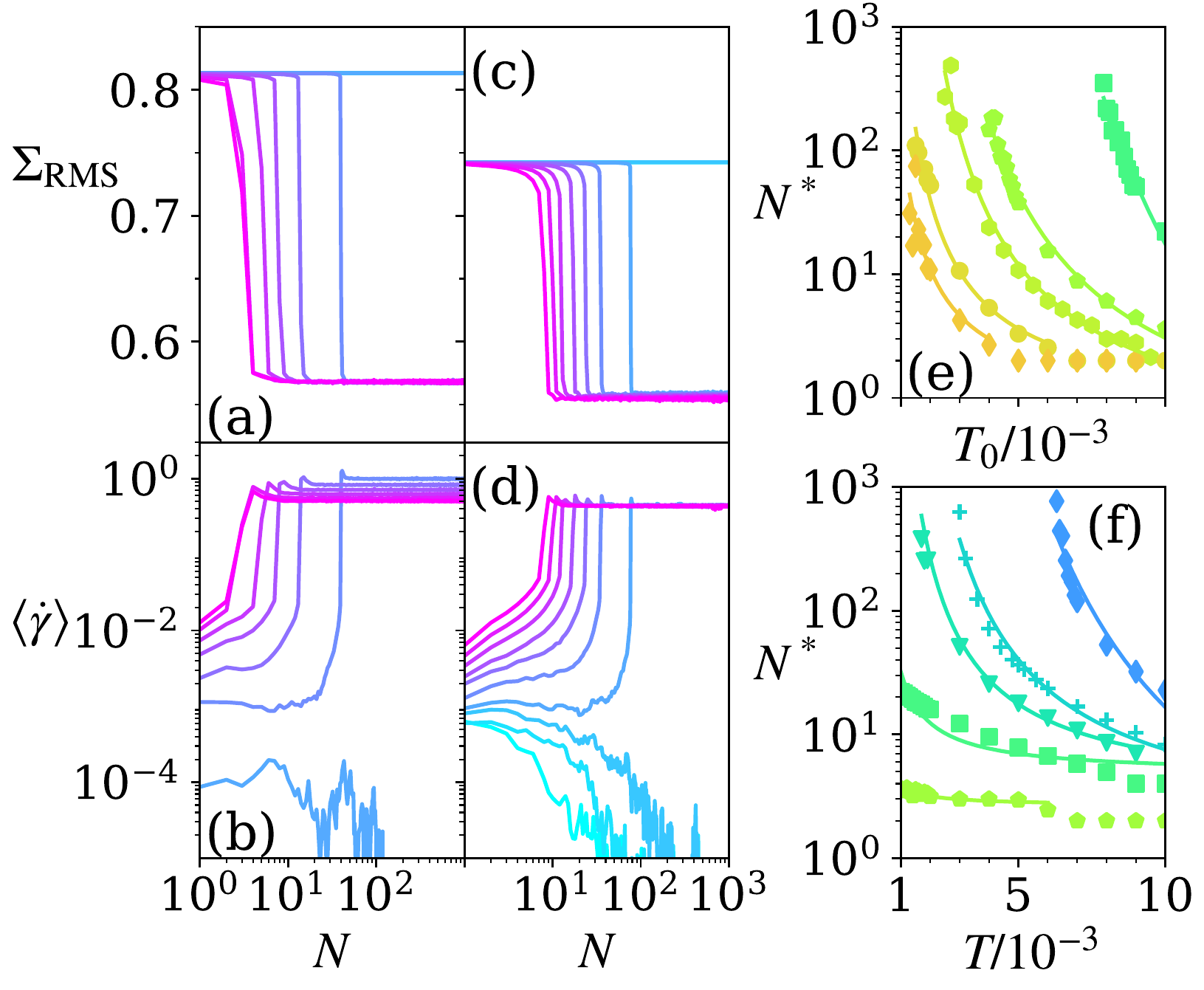}
    \caption{{\bf TEP model.} {\bf a)} Root mean square stress and {\bf b)} mean degree of shear banding over each cycle as a function of cycle number $N$ for pre-quench temperatures $T_0 = 0.001,0.002,...,0.010$ in curves with drops in (a,c) and rises in (b,d) right to left. $\gamma_0 = 1.15$, $T = 0.001$. {\bf c)+d)} Counterpart curves for working temperatures $T = 0.001,0.002,...,0.010$ in curves turquoise to magenta. $\gamma_0 = 1.05$, $T_0 = 0.01$. {\bf e)} Cycle number at yielding $N^*$ vs. pre-quench temperature $T_0$ for strain amplitudes $\gamma_0 = 1.10, 1.15, 1.17, 1.20, 1.22$ in curves downward. $T = 0.001$. Solid lines: fits to $N^* = Ae^{B/T_0}$.  {\bf f)} $N^*$ vs. working temperature $T$  for $\gamma_0 = 1.00, 1.05, 1.07, 1.10, 1.15$ in curves downward.  $T_0 = 0.01$. Solid lines: fits to $N^* = Ce^{D/T}$.
    }
    \label{fig:TEP2}
\end{figure}

{\it TEP results ---} We now show the same physics to obtain in the TEP model, thereby increasing confidence that it will be generic across many amorphous materials. Figs.~\ref{fig:TEP1}a,b) and~\ref{fig:TEP2}a-d) again show a two-stage yielding process, with strain heterogeneity slowly accumulating and  the stress barely declining, before catastrophic failure in which the stress suddenly drops and shear bands form.  The number of cycles $N^*$ before failure again increases dramatically with decreasing imposed strain $\gamma_0$, as seen for several pre-quench temperatures $T_0$ in Fig.~\ref{fig:TEP1}c) and working temperatures $T$ in d).  An interesting difference between TEP and SGR is also apparent. In SGR, recall that $N^*$ increases rapidly with decreasing $\gamma_0$, but with no  apparent divergence before the magnitude of yielding becomes negligible (Fig.~\ref{fig:SGRgamma}, c-e). In TEP, $N^*$ diverges at a non-zero $\gamma_0$ for which yielding is still strongly apparent (Fig.~\ref{fig:TEP1}c+d). Whether this constitutes a fundamental difference between the  models or is simply due to our TEP results being for lower $T$ and stronger annealing than are computationally accessible in SGR is unclear.

We now consider the way in which yielding depends in TEP on the degree to which the sample is annealed prior to shear.  In Fig.~\ref{fig:TEP2}a,b), a collection of yielding curves for decreasing annealing temperature $T_0$ in curves left to right demonstrates a dramatically increasing delay before yielding with increasing sample annealing (lower $T_0$). The number of cycles before yielding is fit to the Boltzmann form $N^* =A\exp(B/T_0)$ in Fig.~\ref{fig:TEP2}e). Ultra-annealed samples ($T_0\to 0$) are thus predicted in TEP to show indefinitely delayed yielding $N^*\to\infty$, in close analogy with the corresponding limit $\tw\to\infty$ in SGR.

We explore finally  the dependence of yielding on working temperature $T$ in TEP. A collection of yielding curves left to right in Fig.~\ref{fig:TEP2}c,d) shows a dramatically increasing delay before yielding with decreasing  $T$. The number of cycles before yielding is fit to the Boltzmann form $N^* =A\exp(B/T)$ in Fig.~\ref{fig:TEP2}f). Accordingly, then, TEP predicts infinitely delayed yielding in the athermal limit of zero working temperature $T\to 0$, at  fixed strain amplitude $\gamma_0$ and pre-quench temperature $T_0$.

{\it Conclusions ---} We have shown the yielding of amorphous materials in oscillatory shear to comprise a two stage process. The first is one of slow fatigue, in which low levels of strain heterogeneity gradually accumulate from cycle to cycle. In the second, the stress drops precipitously and the strain strongly localises into shear bands, leading to catastrophic material failure. The number of cycles $N^*$ before failure increases dramatically with decreasing imposed strain amplitude and increasing  annealing. Finally, $N^*$ diverges in the limit of zero working temperature $T\to 0$, showing that a small non-zero temperature is indispensable to ultra-delayed yielding. 

In future, it would be interesting to consider how the slow fatigue and catastrophic failure studied here (``inter-cycle yielding", over many cycles) relates to  the alternating ``intra-cycle" yielding (with shear banding formation) and resolidification (with rehealing to homogeneous shear) that arises in yield stress fluids once a state has been attained that is invariant from cycle to cycle~\cite{radhakrishnan2016shear,radhakrishnan2018shear,rogers2011sequence,rogers2012sequence}. Another important challenge is to reconcile our divergent $N^*$ in the athermal limit $T\to 0$ with a finite $N^*$ at $T=0$ in the mean field study of Ref.~\cite{parley2022mean}, which neglects banding. It would also be interesting to model yielding in oscillatory shear {\rm stress}, as studied experimentally~\cite{perge2014time,gibaud2016multiple,saint2017predicting}. Indeed, any fundamental similarities and differences between delayed yielding in oscillatory shear and other protocols such as creep should also be considered.  A fuller exploration of the distinction between ductile and brittle yielding is also warranted~\cite{barlow2020ductile}.  

Our predictions  are directly testable experimentally. Bulk rheological measurements of the cycle-to-cycle stress  can be compared  with Figs.~\ref{fig:SGRgamma}a,d),~\ref{fig:SGRtw}a,d)~\ref{fig:phase-space}b),~\ref{fig:TEP1}a) and~\ref{fig:TEP2}a,c). From these stress measurements, the number of cycles to failure $N^*$ can be extracted and compared with  Figs.~\ref{fig:SGRgamma}c),~\ref{fig:SGRtw}c)~\ref{fig:phase-space}a),~\ref{fig:TEP1}c,d) and~\ref{fig:TEP2}e,f).  Ultrasound imaging can be used to measure the velocity field~\cite{manneville2008recent}, from which the cycle-to-cycle degree of shear banding $\Delta\gdot$ can be extracted as prescribed on p2 and compared with our Figs.~\ref{fig:SGRgamma}b,f),~\ref{fig:SGRtw}b,f),~\ref{fig:TEP1}b) and~\ref{fig:TEP2}b,d). All these quantities can also be accessed directly in direct particle simulations.

{\it Acknowledgements ---} We thank Jack Parley and Peter Sollich for interesting discussions. This project has received funding from the European Research Council (ERC) under the European Union's Horizon 2020 research and innovation programme (grant agreement No. 885146).  J.O.C was supported by the EPSRC funded Centre for Doctoral Training in Soft Matter and Functional Interfaces (SOFI CDT - EP/L015536/1).


%

\end{document}